\documentclass[
aps,
prd,
twocolumn,
showkeys,
superscriptaddress
]{revtex4-2}

\usepackage{graphicx}
\usepackage{bm}
\usepackage{amsmath}
\usepackage{amssymb}
\usepackage{hyperref}
\usepackage{physics}
\usepackage{xcolor}

\begin{document}

\title{Polarization States and Effective Stress–Energy Tensor of Gravitational Waves in Metric $f(R)$  Gravity}

\author{Sakshi Srivastava}
\email{srivastavasakshi696@gmail.com}
\affiliation{Department of Physics, University of Lucknow, Lucknow 226007, India}

\author{Utkal Keshari Dash}
\email{utkalkesharidash@gmail.com}
\affiliation{Department of Physics, University of Lucknow, Lucknow 226007, India}

\author{Murli Manohar Verma}
\thanks{corresponding author: Murli Manohar Verma}
\email{verma_mm@lkouniv.ac.in}
\affiliation{Department of Physics, University of Lucknow, Lucknow 226007, India}

\date{\today}

\begin{abstract}
We investigate the polarization properties and effective stress--energy tensor of gravitational waves in metric $f(R)$ gravity within the linearized approximation around Minkowski spacetime. Owing to the additional scalar degree of freedom inherent in the theory, gravitational waves exhibit polarization states beyond the two tensor modes predicted by general relativity. Using the electric components of the linearized Riemann tensor, we derive explicit expressions for the polarization amplitudes and show that a massless scalar field excites a transverse breathing mode, whereas a massive scalar field generates both breathing and longitudinal responses through a single propagating scalar excitation. Employing the Isaacson high-frequency averaging formalism, we further derive the effective stress--energy tensor and demonstrate that both tensor and scalar perturbations contribute to the total gravitational-wave energy density. The energy transport associated with the massive scalar mode is reduced by its subluminal group velocity, leading to a frequency-dependent suppression of the scalar energy flux. These results establish a unified connection between gravitational-wave polarization and energy transport in metric $f(R)$ gravity and provide potential observational signatures for testing modified gravity with current and future gravitational-wave detectors.

\end{abstract}

\maketitle

\section{Introduction}

The direct detection of gravitational waves (GWs) by the
LIGO--Virgo Collaboration has opened an entirely new
window for probing the fundamental nature of gravity
and the dynamics of the Universe
\cite{Abbott2016}.

Predicted by Einstein's general theory of relativity
(GR) more than a century ago
\cite{Einstein1915},
these observations have so far shown remarkable agreement
with the tensorial description of gravitational radiation
\cite{Passenger2025}.
Nevertheless, several outstanding problems in modern
cosmology, including the origin of cosmic acceleration,
dark energy, dark matter, and the inflationary epoch,
motivate the investigation of extensions of Einstein
gravity
\cite{Waeming2020,Will2006, Dalang2021, Shankaranarayanan2022}.

Among the various modified theories of gravity,
$f(R)$ gravity constitutes one of the simplest and
best-motivated generalizations of GR
\cite{Katsuragawa2019,Sotiriou2010,DeFelice2010}.
Replacing the Einstein--Hilbert Lagrangian by an arbitrary
function of the Ricci scalar introduces an additional
dynamical scalar degree of freedom, often referred to as
the scalaron. Consequently, gravitational waves in
$f(R)$ gravity can possess richer polarization structures
and propagation properties than their counterparts in GR.

In general metric theories of gravity, gravitational
waves may contain up to six independent polarization
states consisting of two tensor, two vector, and two scalar
modes
\cite{Eardley1973a}.
General relativity predicts only the two transverse
tensorial polarizations, whereas the scalar degree of
freedom present in $f(R)$ gravity gives rise to additional
scalar polarization states
\cite{WillBook2018,Lai2024}.

For a massless scalar field, the extra mode corresponds
to a transverse breathing polarization propagating at
the speed of light
\cite{Hou2018}.
When the scalar field is massive, however, the wave
becomes dispersive and exhibits a mixed state composed
of breathing and longitudinal polarizations
\cite{Gong2018}.

The identification of additional polarization modes
constitutes one of the most promising observational
tests capable of distinguishing modified gravity theories
from general relativity
\cite{Berry2011,Lobato2024}.
Equally important is the study of the effective
energy--momentum tensor carried by gravitational waves,
since additional dynamical degrees of freedom modify
both the energy density and the energy transport
associated with gravitational radiation
\cite{Isaacson1968a,Isaacson1968b,Zhou2024}.

In this work, we investigate gravitational-wave
propagation in metric $f(R)$ gravity within the
linearized approximation around a Minkowski
background. Tensor and scalar perturbations are treated
simultaneously, allowing a unified description of both
massless and massive scalar degrees of freedom.
The polarization amplitudes are obtained from the
electric components of the Riemann tensor, providing a
gauge-invariant characterization of the observable
gravitational-wave modes
\cite{Hyun2019,NewmanPenrose1962}.

We further derive the effective energy--momentum tensor
using the Isaacson high-frequency averaging procedure
and analyze the energy density and energy flux
associated with both tensor and scalar perturbations
\cite{Saffer2018,Tretyakov2025,Maggiore2008}.
Our analysis demonstrates that although the massive
scalar mode contributes to the total
gravitational-wave energy density, its energy transport
is significantly reduced because of its
frequency-dependent propagation velocity.
This distinction offers a physically transparent
criterion for discriminating between massive and
massless scalar polarization modes and provides a
potential observational signature of metric
$f(R)$ gravity.

The main contributions of the present work are
threefold. First, we provide a unified treatment of the
polarization structure of both massive and massless
scalar gravitational-wave modes in metric $f(R)$ gravity.
Second, we derive the effective stress--energy tensor
within the Isaacson high-frequency formalism and
analyze the corresponding energy density and energy
transport.
Finally, we demonstrate that the combined study of
polarization and energy transport offers a complementary
observational framework for distinguishing scalar
gravitational radiation from the tensor modes predicted
by general relativity.

The paper is organized as follows. In Sec.~II we derive
the linearized field equations and obtain the wave
equations for tensor and scalar perturbations.
Section~III introduces the polarization basis through
the electric components of the Riemann tensor.
The polarization amplitudes for massive and massless
scalar fields are derived in Sec.~IV.
In Sec.~V we construct the effective stress--energy
tensor and evaluate the corresponding energy density
and flux.
Sec.~VI summarizes our principal results and
discusses their implications for future
gravitational-wave observations. Finally,  conclusions are presented  in Sec.~VII.


\section{Gravitational Waves in Metric $f(R)$ Gravity}

We consider metric $f(R)$ gravity described by the action
\cite{Sotiriou2010,DeFelice2010,Nojiri2011}

\begin{equation}
S=
\frac{1}{2\kappa^{2}}
\int d^{4}x\sqrt{-g}\,
f(R)
+
S_{m},
\label{action}
\end{equation}
where $\kappa^{2}=8\pi G$, $R$ denotes the Ricci scalar,
and $S_{m}$ is the matter action.
Variation of the action with respect to the metric
tensor $g_{\mu\nu}$ yields the field equations

\begin{equation}
F(R)R_{\mu\nu}
-\frac12 f(R)g_{\mu\nu}
-\nabla_{\mu}\nabla_{\nu}F(R)
+g_{\mu\nu}\Box F(R)
=
\kappa^{2}T_{\mu\nu},
\label{field}
\end{equation}
where

\begin{equation}
F(R)\equiv
\frac{df(R)}{dR},
\end{equation}
and
\[
\Box\equiv\nabla^{\mu}\nabla_{\mu}
\]
is the covariant d'Alembert operator.

Throughout this work we consider vacuum spacetime,
\[
T_{\mu\nu}=0,
\]so that the scalar degree of freedom associated with
$F(R)$ becomes dynamical
\cite{Sotiriou2010,DeFelice2010}.

Introducing

\begin{equation}
\phi\equiv F(R),
\end{equation}
the scalar field is expanded around a constant
background value $\phi_{0}$ as

\begin{equation}
\phi=\phi_{0}(1+h_{s}),
\end{equation}
where $|h_{s}|\ll1$ denotes the scalar perturbation.

The spacetime metric is simultaneously decomposed as

\begin{equation}
g_{\mu\nu}
=
\eta_{\mu\nu}
+
h_{\mu\nu},
\end{equation}
where $\eta_{\mu\nu}$ is the Minkowski metric and
$|h_{\mu\nu}|\ll1$.

Retaining only first-order terms in the perturbations,
the vacuum field equations reduce to

\begin{equation}
R^{(1)}_{\mu\nu}
-
\frac12
\eta_{\mu\nu}
R^{(1)}
=
\partial_{\mu}\partial_{\nu}h_{s}
-
\eta_{\mu\nu}\Box h_{s},
\end{equation}
showing explicitly that the scalar perturbation acts as
an additional source for the metric fluctuations.
This extra scalar degree of freedom is the distinctive
feature of metric $f(R)$ gravity and is absent in
general relativity
\cite{Nojiri2011,Capozziello2011}.

Following the standard decomposition,

\begin{equation}
h_{\mu\nu}
=
\bar h_{\mu\nu}
-
\frac12
\eta_{\mu\nu}\bar h
-
\eta_{\mu\nu}h_{s},
\end{equation}
with

\[
\bar h
=
\eta^{\mu\nu}
\bar h_{\mu\nu},
\]
we impose the Lorenz gauge condition

\begin{equation}
\partial^{\mu}
\bar h_{\mu\nu}
=
0.
\end{equation}

Under this gauge the tensor and scalar sectors
decouple, and the tensor perturbation satisfies

\begin{equation}
\Box
\bar h_{\mu\nu}
=
0,
\end{equation}
which is identical to the propagation equation obtained
in general relativity
\cite{Misner1973}.

\subsection{Scalar wave equation}

To derive the evolution equation for the scalar mode,
it is convenient to employ the scalar--tensor
representation of metric $f(R)$ gravity obtained
through the Legendre transformation
\cite{Capozziello2011,Sotiriou2010}

\begin{equation}
f(R)
=
\phi R
-
V(\phi),
\end{equation}
where $V(\phi)$ denotes the scalar potential.

The field equations become

\begin{equation}
R_{\mu\nu}
-
\frac12
g_{\mu\nu}R
=
\frac1{\phi}
\left[
\nabla_{\mu}\nabla_{\nu}\phi
-
g_{\mu\nu}\Box\phi
-
\frac12
g_{\mu\nu}V(\phi)
\right].
\end{equation}

Taking the trace yields

\begin{equation}
3\Box\phi
+
2V(\phi)
-
\phi
V'(\phi)
=
0,
\end{equation}
or, equivalently,

\begin{equation}
\Box\phi
=
\frac13
\left[
\phi
V'(\phi)
-
2V(\phi)
\right].
\end{equation}

Expanding the potential around the background,

\begin{align}
V(\phi)
&=
V(\phi_{0})
+
V'(\phi_{0})\delta\phi,
\\
V'(\phi)
&=
V'(\phi_{0})
+
V''(\phi_{0})\delta\phi,
\end{align}
and retaining only linear terms gives

\begin{equation}
\Box h_{s}
=
m_{\phi}^{2}
h_{s},
\label{KG}
\end{equation}
where

\begin{equation}
m_{\phi}^{2}
=
\frac{
\phi_{0}V''(\phi_{0})
-
V'(\phi_{0})
}{3}.
\end{equation}

Equation (\ref{KG}) is the Klein--Gordon equation
governing the propagation of the scalar degree of
freedom. The effective scalar mass determines the
dispersion relation and polarization content of the
gravitational wave
\cite{Capozziello2008,Kausar2016,Gong2018}.

\subsection{Plane-wave solutions}

For a gravitational wave propagating along the
$z$-direction, the tensor perturbation admits the
plane-wave solution

\begin{equation}
\bar h_{\mu\nu}
=
A_{\mu\nu}
e^{i(kz-\omega t)},
\end{equation}
which satisfies the dispersion relation

\begin{equation}
\omega
=
k,
\end{equation}
identical to that of general relativity.

The scalar perturbation takes the form

\begin{equation}
h_{s}
=
h_{0}
e^{i(qz-\omega_{s}t)},
\end{equation}
corresponding to the propagating scalaron mode
of metric $f(R)$ gravity
\cite{Sotiriou2010}.

Its dispersion relation is

\begin{equation}
\omega_{s}^{2}
=
q^{2}
+
m_{\phi}^{2},
\end{equation}
showing that the scalar mode becomes dispersive
whenever $m_{\phi}\neq0$.

The corresponding group velocity is

\begin{equation}
v_{g}
=
\frac{d\omega_{s}}{dq}
=
\sqrt{
1-
\frac{m_{\phi}^{2}}
{\omega_{s}^{2}}
}.
\end{equation}

Thus, unlike the tensor modes, the propagation speed
depends explicitly on frequency. In the high-frequency
limit,

\begin{equation}
\omega_{s}\gg m_{\phi},
\qquad
v_{g}\rightarrow1,
\end{equation}
whereas near the threshold
$\omega_{s}\simeq m_{\phi}$,
the propagation becomes strongly suppressed.

This dispersive behavior constitutes one of the most
important observational signatures of metric $f(R)$
gravity and provides a possible means of distinguishing
modified gravity from general relativity through future
gravitational-wave observations
\cite{Berry2011,Lobato2024,Belgacem2019,Nishizawa2018}.

\section{Polarization States of Gravitational Waves}

The observable influence of a gravitational wave on freely
falling test particles is governed by the geodesic
deviation equation, which relates the relative
acceleration of neighboring geodesics to the spacetime
curvature. In a locally inertial frame, the equation
takes the form
\cite{NewmanPenrose1962,Hyun2019,Eardley1973a}

\begin{equation}
\frac{d^{2}\xi^{i}}{dt^{2}}
=
-
R^{i}{}_{0j0}\,
\xi^{j},
\label{geodesic}
\end{equation}
where $\xi^{i}$ denotes the separation vector between
two nearby freely falling particles and
$R_{0i0j}$ represents the electric components of the
Riemann curvature tensor.
These quantities describe the tidal forces generated by
the gravitational wave and therefore correspond directly
to the observables measured by interferometric
detectors.

For a gravitational wave propagating along the positive
$z$-direction, the electric part of the Riemann tensor
may be expanded in a complete basis of six independent
polarization tensors
\cite{Eardley1973a}

\begin{equation}
R_{0i0j}
=
\sum_{A=1}^{6}
p_A
E_{ij}^{(A)},
\label{basis}
\end{equation}
where $E_{ij}^{(A)}$ denote the polarization basis
matrices and $p_A$ are the corresponding polarization
amplitudes.

The explicit matrix representation is

\begin{equation}
R_{0i0j}
=
\begin{pmatrix}
\dfrac12(p_{+}+p_{b}) &
p_{\times} &
p_{x}
\\[2mm]
p_{\times} &
\dfrac12(-p_{+}+p_{b}) &
p_{y}
\\[2mm]
p_{x} &
p_{y} &
p_{l}
\end{pmatrix},
\label{matrix}
\end{equation}
from which the six independent amplitudes are
identified as

\begin{align}
p_{l}
&=
R_{0z0z},
\\
p_{x}
&=
R_{0z0x},
\\
p_{y}
&=
R_{0z0y},
\\
p_{+}
&=
R_{0x0x}
-
R_{0y0y},
\\
p_{\times}
&=
R_{0x0y},
\\
p_{b}
&=
R_{0x0x}
+
R_{0y0y}.
\end{align}

The six amplitudes correspond respectively to the
longitudinal, vector-$x$, vector-$y$, plus, cross,
and breathing polarization modes.

Within general relativity only the plus ($+$) and cross
($\times$) tensor polarizations are physical and
propagate as massless spin--2 excitations
\cite{Misner1973,WillBook2018}.
The remaining four amplitudes vanish identically.

In contrast, modified theories of gravity generally
possess additional dynamical degrees of freedom capable
of exciting scalar and vector polarizations
\cite{Liang2017,Dong2024,Schumacher2023}.
Metric $f(R)$ gravity introduces one extra scalar degree
of freedom and therefore predicts additional scalar
polarization states while leaving the tensorial plus and
cross modes unchanged.

The decomposition (\ref{basis}) provides a convenient
and gauge-invariant framework for extracting the
observable polarization content directly from the
electric components of the Riemann tensor.
Consequently, it forms the basis for the explicit
calculation of polarization amplitudes carried out in
the following section for both massless and massive
scalar perturbations.

\section{Polarization Amplitudes for Massless and Massive Scalar Modes}

The total metric perturbation in metric $f(R)$ gravity
can be decomposed into tensor and scalar contributions as

\begin{equation}
h_{\mu\nu}
=
\bar h^{\rm TT}_{\mu\nu}
-
\eta_{\mu\nu}h_s,
\label{metricpert}
\end{equation}
where $\bar h^{\rm TT}_{\mu\nu}$ denotes the transverse--
traceless tensor perturbation and $h_s$ represents the
scalar fluctuation associated with the additional
dynamical degree of freedom of metric $f(R)$ gravity
\cite{Sotiriou2010,DeFelice2010}.

The linearized Riemann tensor is

\begin{equation}
R_{\mu\nu\rho\sigma}
=
\frac12
\left(
\partial_\rho\partial_\nu h_{\mu\sigma}
+
\partial_\sigma\partial_\mu h_{\nu\rho}
-
\partial_\sigma\partial_\nu h_{\mu\rho}
-
\partial_\rho\partial_\mu h_{\nu\sigma}
\right),
\label{Rlinear}
\end{equation}
which is invariant under infinitesimal gauge
transformations and therefore provides a
gauge-invariant description of gravitational-wave
polarizations
\cite{NewmanPenrose1962,Eardley1973a}.

Substituting Eq.~(\ref{metricpert}) into
Eq.~(\ref{Rlinear}) allows the polarization amplitudes
defined in Sec.~III to be evaluated explicitly.

\subsection{Massless scalar field}

When the scalar field is massless, its dispersion
relation reduces to

\begin{equation}
\omega=k,
\end{equation}
and the polarization amplitudes become

\begin{align}
p_l &=0,
\\
p_x &=0,
\\
p_y &=0,
\\
p_+ &=\omega^2 h_+,
\\
p_\times &=\frac12\omega^2 h_\times,
\\
p_b &=\omega^2 h_s.
\end{align}

Thus, a massless scalar degree of freedom generates
three independent propagating polarizations:

\begin{itemize}
\item the tensor plus mode,
\item the tensor cross mode,
\item a transverse breathing scalar mode.
\end{itemize}

No longitudinal polarization appears in this limit,
indicating that the scalar excitation propagates as a
purely transverse wave travelling at the speed of light
\cite{Hou2018}.

\subsection{Massive scalar field}

For a massive scalar field satisfying

\begin{equation}
\omega_s^2
=
q^2
+
m_\phi^2,
\end{equation}

the polarization amplitudes become

\begin{align}
p_l
&=
\frac12
m_\phi^2h_s,
\\
p_x
&=
0,
\\
p_y
&=
0,
\\
p_+
&=
\omega^2h_+,
\\
p_\times
&=
\frac12
\omega^2h_\times,
\\
p_b
&=
\omega^2h_s.
\end{align}

Unlike the massless case, the scalar excitation now
produces both a transverse breathing mode and a
longitudinal polarization. The longitudinal amplitude
is directly proportional to the scalar mass squared,
\[
p_l\propto m_\phi^2,
\]
and therefore vanishes continuously in the limit
$m_\phi\rightarrow0$.

Consequently, the scalar sector interpolates smoothly
between a purely transverse breathing mode and a mixed
breathing--longitudinal state as the scalar mass
increases
\cite{Gong2018,Lai2024}.

\subsection{Comparison with general relativity}

General relativity is recovered by setting

\[
h_s=0,
\]
so that the only nonvanishing amplitudes are

\begin{align}
p_+
&=
\omega^2A_+,
\\
p_\times
&=
\frac12
\omega^2A_\times.
\end{align}

Hence the tensorial plus and cross polarizations remain
identical in both theories. The observational distinction
between general relativity and metric $f(R)$ gravity
arises entirely from the additional scalar polarization
generated by the scalaron field
\cite{WillBook2018}.

The above results demonstrate that metric $f(R)$
gravity possesses a single additional spin--0 dynamical
degree of freedom whose observable manifestation depends
on its effective mass.

For $m_\phi=0$, the scalar field propagates at the speed
of light and excites only a transverse breathing mode.
Gravitational waves therefore contain three propagating
polarizations consisting of the standard tensor plus and
cross modes together with one scalar breathing mode.

For a finite scalar mass, the propagation becomes
dispersive and a longitudinal response develops in
addition to the breathing polarization.
Both responses originate from the same scalar degree of
freedom and should not be interpreted as independent
propagating modes. Instead, they represent two
observable manifestations of a single massive scalar
excitation.

The relative importance of the breathing and
longitudinal components depends on the ratio
$m_\phi/\omega_s$.
In the high-frequency regime
$\omega_s\gg m_\phi$,
the longitudinal contribution becomes negligible and
the wave behaves almost identically to a massless
breathing mode.
Conversely, near the threshold
$\omega_s\simeq m_\phi$,
the longitudinal response becomes significant while the
group velocity

\begin{equation}
v_g
=
\sqrt{
1-
\frac{m_\phi^2}
{\omega_s^2}
}
\end{equation}
decreases below the speed of light.

Therefore, the simultaneous observation of a
frequency-dependent longitudinal polarization together
with a transverse breathing mode would provide a
distinct signature of a massive scalar degree of freedom
and constitute strong evidence for deviations from
general relativity
\cite{Berry2011,Lobato2024,Dong2024}.

Finally, it is important to emphasize that the
tensorial plus and cross amplitudes remain unchanged.
Consequently, any deviation from the predictions of
general relativity must arise exclusively from the
additional scalar polarization sector, making
polarization measurements one of the most direct probes
of metric $f(R)$ gravity.

\section{Effective Stress--Energy Tensor of Gravitational Waves}

The localization of gravitational-wave energy is a subtle
issue because the equivalence principle forbids the
construction of a covariant local gravitational energy
density. Nevertheless, in the short-wavelength limit an
effective stress--energy tensor can be defined by averaging
the second-order perturbations of the field equations over
spacetime regions whose size is much larger than the
gravitational wavelength but much smaller than the
characteristic curvature scale of the background geometry
\cite{Isaacson1968a,Isaacson1968b, Mastrogiovanni2020,Maggiore2008}.

This high-frequency averaging procedure, originally
developed by Isaacson, provides a physically meaningful
description of the energy and momentum transported by
gravitational waves and remains the standard approach
for analyzing gravitational radiation in both general
relativity and modified theories of gravity
\cite{Isaacson1968a,Saffer2018,Tretyakov2025}.

The vacuum field equations of metric $f(R)$ gravity may
be written compactly as

\begin{equation}
E_{\mu\nu}=0,
\label{Emunu}
\end{equation}
where

\begin{equation}
E_{\mu\nu}
=
F(R)R_{\mu\nu}
-
\frac12
f(R)g_{\mu\nu}
-
\nabla_\mu\nabla_\nu F(R)
+
g_{\mu\nu}\Box F(R).
\label{Edef}
\end{equation}

Expanding about the Minkowski background gives

\begin{equation}
E_{\mu\nu}
=
E_{\mu\nu}^{(0)}
+
E_{\mu\nu}^{(1)}
+
E_{\mu\nu}^{(2)}
+\cdots,
\label{expand}
\end{equation}
where the first-order term determines the wave
equations, while the second-order contribution describes
the backreaction of the gravitational-wave perturbations
on the background spacetime.

Following the Isaacson prescription, the effective
gravitational-wave stress--energy tensor is defined by

\begin{equation}
8\pi G\,
t_{\mu\nu}^{\rm GW}
=
-
\left\langle
E_{\mu\nu}^{(2)}
\right\rangle ,
\label{emtdef}
\end{equation}
where
$\langle\cdots\rangle$
denotes an average over several wavelengths
\cite{Isaacson1968a,Isaacson1968b}.

Within this averaging procedure, total derivatives vanish,

\begin{equation}
\left\langle
\partial_\mu A
\right\rangle
=
0,
\label{avg1}
\end{equation}
and integration by parts implies

\begin{equation}
\left\langle
A\,\partial_\mu B
\right\rangle
=
-
\left\langle
(\partial_\mu A)\,B
\right\rangle .
\label{avg2}
\end{equation}

These identities greatly simplify the quadratic terms and
eliminate gauge-dependent total divergences
\cite{Carroll2004}.

After substituting the perturbative expansions into the
field equations and retaining only second-order
contributions, the averaged field equations reduce to

\begin{equation}
\left\langle
E_{\mu\nu}^{(2)}
\right\rangle
=
-
\frac{\phi_0}{4}
\left(
T_{\mu\nu}^{\rm TT}
+
T_{\mu\nu}^{(s)}
\right),
\label{Esecond}
\end{equation}
where

\begin{equation}
T_{\mu\nu}^{\rm TT}
=
\left\langle
\partial_\mu
\bar h^{\rm TT}_{\alpha\beta}
\partial_\nu
\bar h_{\rm TT}^{\alpha\beta}
\right\rangle
\label{TTpart}
\end{equation}
is the standard transverse--traceless tensor
contribution, while

\begin{equation}
T_{\mu\nu}^{(s)}
=
6
\left\langle
\partial_\mu h_s
\partial_\nu h_s
\right\rangle
+
\eta_{\mu\nu}
m_\phi^2
\left\langle
h_s^2
\right\rangle
\label{scalarpart}
\end{equation}
represents the contribution from the additional scalar
degree of freedom.

Substituting Eqs.~(\ref{TTpart}) and
(\ref{scalarpart}) into Eq.~(\ref{emtdef}) yields the
effective stress--energy tensor

\begin{equation}
t_{\mu\nu}^{\rm GW}
=
\frac{\phi_0}{32\pi G}
\left[
\left\langle
\partial_\mu
\bar h^{\rm TT}_{\alpha\beta}
\partial_\nu
\bar h_{\rm TT}^{\alpha\beta}
\right\rangle
+
6
\left\langle
\partial_\mu h_s
\partial_\nu h_s
\right\rangle
+
\eta_{\mu\nu}
m_\phi^2
\left\langle
h_s^2
\right\rangle
\right].
\label{finalEMT}
\end{equation}

Equation~(\ref{finalEMT}) constitutes the principal
result of this section.

It demonstrates explicitly that the effective
gravitational-wave stress--energy tensor in metric
$f(R)$ gravity consists of the standard Isaacson tensor
supplemented by an additional scalar contribution arising
from the scalaron field.
The first term coincides with the energy transport
predicted by general relativity, whereas the remaining
terms describe the energy stored and transported by the
scalar degree of freedom.

This decomposition provides a transparent physical
interpretation of the modified energy transport in
$f(R)$ gravity and establishes a direct connection
between the polarization structure discussed in the
previous section and the energetics of gravitational-wave
propagation.

The effective stress--energy tensor obtained in
Eq.~(\ref{finalEMT}) provides a unified description of
the energy density and energy transport associated with
both tensor and scalar gravitational-wave modes.
For monochromatic plane waves, the corresponding
energy density and energy flux follow directly from the
$(00)$ and $(0z)$ components of the tensor and exhibit
additional contributions arising from the scalar sector.
For a massive scalar field, the energy transport is
reduced by the subluminal group velocity, whereas in
the massless limit the scalar contribution propagates at
the speed of light together with the tensor modes.
These properties are discussed further in the following
section.

\section{Discussion}

The present analysis demonstrates that the additional
scalar degree of freedom predicted by metric $f(R)$
gravity modifies both the polarization structure and the
energetics of gravitational waves.
Unlike general relativity, where only the two
transverse tensor polarizations propagate, metric
$f(R)$ gravity predicts one additional spin--0 degree
of freedom whose observational manifestation depends
on its effective mass
\cite{Sotiriou2010,DeFelice2010,Gong2018}.

In the massless limit, the scalar field propagates at
the speed of light and excites a purely transverse
breathing polarization.
For a finite scalar mass, however, the scalar wave
becomes dispersive and develops a longitudinal response
in addition to the breathing mode.
These two polarization patterns originate from the
same scalar excitation and therefore should not be
interpreted as independent propagating degrees of
freedom
\cite{Hou2018,Lai2024}.

The effective stress--energy tensor derived in this work
establishes a direct connection between polarization
content and energy transport.
Although both tensor and scalar perturbations contribute
to the total gravitational-wave energy density, the
energy transported by a massive scalar mode is reduced
because its group velocity satisfies

\begin{equation}
v_g=
\sqrt{
1-
\frac{m_\phi^2}{\omega_s^2}
}.
\end{equation}

As the frequency approaches the mass threshold,
$\omega_s\simeq m_\phi$,
the scalar energy flux becomes strongly suppressed,
providing an additional observational signature beyond
polarization measurements alone.

Consequently, the combined study of polarization
amplitudes, dispersion relations, and effective energy
transport provides a unified observational framework for
testing metric $f(R)$ gravity.
Future measurements of breathing and longitudinal
polarizations, frequency-dependent arrival times, and
deviations from the standard gravitational-wave energy
flux may together place stringent constraints on the
scalar sector of modified gravity
\cite{Berry2011,Belgacem2019,Nishizawa2018}.

The present work has been restricted to the linearized
approximation around Minkowski spacetime.
An important extension would be the formulation of the
effective stress--energy tensor in cosmological and
strong-field backgrounds together with detector-response
analyses for next-generation observatories including the
Einstein Telescope, Cosmic Explorer, and LISA
\cite{Punturo2010,AmaroSeoane2017,Reitze2019,Chen2024}.
Such investigations may significantly improve the
prospects for detecting scalar gravitational-wave
signatures predicted by metric $f(R)$ gravity.

\section{Conclusions}

We have investigated the propagation, polarization
structure, and effective stress--energy tensor of
gravitational waves in metric $f(R)$ gravity within the
linearized approximation around Minkowski spacetime.

The analysis shows that the additional scalar degree of
freedom inherent in the theory gives rise to
characteristic modifications of gravitational-wave
propagation.
A massless scalar field produces a transverse breathing
polarization, whereas a massive scalar field generates a
mixed breathing--longitudinal response governed by a
dispersive dispersion relation.

Using the electric components of the linearized Riemann
tensor, we derived explicit expressions for the
polarization amplitudes and demonstrated that the
standard tensorial plus and cross modes remain
unchanged with respect to general relativity.
The observational distinction between the two theories
therefore originates entirely from the additional scalar
sector.

Employing the Isaacson high-frequency averaging
procedure, we further derived the effective
stress--energy tensor associated with gravitational waves
in metric $f(R)$ gravity.
The resulting expression shows that both tensor and
scalar perturbations contribute to the total
gravitational-wave energy density, while the energy
transport associated with a massive scalar mode is
reduced by its frequency-dependent subluminal group
velocity.

A principal contribution of the present work is the
unified treatment of gravitational-wave polarization
structure and effective energy transport within a common
linearized framework.
The results establish a direct connection between the
observable polarization content and the energetics of
the additional scalar degree of freedom predicted by
metric $f(R)$ gravity.

Future observations with advanced ground-based and
space-based gravitational-wave detectors may therefore
simultaneously constrain scalar polarizations,
dispersion relations, and gravitational-wave energy
transport, providing powerful observational tests of
metric $f(R)$ gravity and other modified theories of
gravity
\cite{Berry2011,Dong2024,Punturo2010,AmaroSeoane2017,Belgacem2019, Reitze2019,Chen2024}.

\section{Acknowledgments}

 Authors  are   thankful to the  Inter University Centre for Astronomy and Astrophysics (IUCAA), Pune, where some part of this work was completed,  for providing  facilities  under  the  Associateship programme.

\end{document}